%% file: main.tex
\documentclass[acmsmall]{acmart}
\AtBeginDocument{%
  }

\input{sec/macro}

\begin{document}


\title{\textsc{ExpeRepair}: Dual-Memory Enhanced LLM-Based Repository-Level Program Repair}

\author{Fangwen Mu}
\authornote{Also With University of Chinese Academy of Sciences}
\email{fangwen2020@iscas.ac.cn}
\affiliation{%
\institution{State Key Laboratory of Intelligent Game, Institute of Software, CAS}
\city{Beijing}
\country{China}
}

\author{Junjie Wang}
\authornotemark[1]
\affiliation{%
\institution{State Key Laboratory of Intelligent Game, Institute of Software, CAS}
\city{Beijing}
\country{China}}
\email{junjie@iscas.ac.cn}

\author{Lin Shi}
\affiliation{%
\institution{Beihang University}
\city{Beijing}
\country{China}}
\email{shilin@buaa.edu.cn}

\author{Song Wang}
\affiliation{%
\institution{York University}
\city{Toronto}
\country{Canada}}
\email{wangsong@yorku.ca}

\author{Shoubin Li}
\authornotemark[1]
\affiliation{%
\institution{State Key Laboratory of Intelligent Game, Institute of Software, CAS}
\city{Beijing}
\country{China}}
\email{shoubin@iscas.ac.cn}

\author{Qing Wang}
\authornotemark[1]
\authornote{Corresponding author}
\affiliation{%
\institution{State Key Laboratory of Intelligent Game, Institute of Software, CAS}
\city{Beijing}
\country{China}}
\email{wq@iscas.ac.cn}


\begin{abstract}

Automatically repairing software issues remains a fundamental challenge at the intersection of software engineering and AI. Although recent advances in Large Language Models (LLMs) have demonstrated potential for repository-level repair tasks, current methods exhibit two notable limitations: (1) they often address issues in isolation, neglecting to incorporate insights from previously resolved issues, and (2) they rely on static, rigid prompting strategies that constrain their ability to generalize across diverse and evolving contexts.
We propose \textsc{ExpeRepair}, a novel LLM-based program repair framework inspired by the dual-memory systems of human cognition, where episodic and semantic memory synergistically support learning and decision-making. Unlike existing methods, \textsc{ExpeRepair} continuously learns from historical repair experiences via dual-channel knowledge accumulation, enabling it to adaptively reuse past knowledge during inference.
Specifically, \textsc{ExpeRepair} organizes prior repair knowledge into two complementary memories: an episodic memory that stores concrete repair demonstrations, and a semantic memory that encodes abstract, reflective insights. At inference time, \textsc{ExpeRepair} activates both memory systems by retrieving relevant demonstrations from episodic memory and recalling high-level repair insights from semantic memory. It further enhances adaptability through dynamic prompt composition, integrating both memory types to replace static prompts with context-aware, experience-driven prompts.
We evaluate \textsc{ExpeRepair} on two benchmarks: SWE-Bench Lite and SWE-Bench Verified. Experimental results show that \textsc{ExpeRepair} achieves pass@1 scores of 60.3\% and 74.6\% on the two benchmarks, respectively, achieving the best performance among the evaluated open-source methods. We have open-sourced \textsc{ExpeRepair} at \texttt{https://github.com/ExpeRepair/ExpeRepair}.

\end{abstract}

\ccsdesc[500]{Software and its engineering~Software creation and management}
\ccsdesc[500]{Software and its engineering~Software development techniques}

\keywords{Automated Program Repair, Large Language Model, AI Agents}

\setcopyright{cc}
\setcctype{by}
\acmDOI{10.1145/3808181}
\acmYear{2026}
\acmJournal{PACMSE}
\acmVolume{3}
\acmNumber{FSE}
\acmArticle{FSE174}
\acmMonth{7}
\acmSubmissionID{fse26mainb-p1572-p}
\received{2025-09-12}
\received[accepted]{2026-03-24}


\maketitle

\input{sec/intro}

\input{sec/method}

\input{sec/exp}
\input{sec/results}
\input{sec/discussion}
\input{sec/related_work}
\input{sec/conclusion}

\bibliographystyle{ACM-Reference-Format}
\bibliography{acmart}
\end{document}

%% file: sec/macro.tex
\newcommand{\tool}{\textsc{ExpeRepair}}
\usepackage{graphicx}
\usepackage{subfiles}
\usepackage{array}
\usepackage{amsmath}
\usepackage{multirow}
\usepackage{caption}
\usepackage{diagbox}
\usepackage{subcaption}
\usepackage{colortbl}
 \usepackage{enumitem}
\usepackage{booktabs}
\usepackage{wrapfig}
\usepackage{fancyhdr}
\usepackage[framemethod=TikZ]{mdframed}
\usepackage{tcolorbox}
\makeatletter
\newcommand{\mybox}[1]{%
  \setbox0=\hbox{#1}%
  \setlength{\@tempdima}{\dimexpr\wd0+13pt}%
  \begin{tcolorbox}[boxrule=0.5pt, colback=gray!10, arc=4pt,
      left=6pt,right=6pt,top=6pt,bottom=6pt,boxsep=0pt]
    #1
  \end{tcolorbox}
}

\tcbuselibrary{listings, skins, breakable}

\newtcolorbox{promptbox}[1][]{
  enhanced,
  breakable,
  colback=gray!5,        
  colframe=black,        
  coltitle=white,        
  colbacktitle=black,    
  title={#1},
  fonttitle=\bfseries,
  boxrule=0.5pt,
  arc=2mm,
  left=2mm,
  right=2mm,
  top=1mm,
  bottom=1mm
}

\usepackage{makecell}
\usepackage{float}

\usepackage{listings}

\definecolor{codegreen}{rgb}{0,0.6,0}
\definecolor{codegray}{rgb}{0.5,0.5,0.5}
\definecolor{codepurple}{rgb}{0.58,0,0.82}
\definecolor{backcolour}{rgb}{0.95,0.95,0.92}

\lstdefinestyle{mystyle}{
  language=Java,
  aboveskip=3mm,
  showstringspaces=false,
  columns=flexible,
  numbers=none,
  backgroundcolor=\color{backcolour},
  commentstyle=\color{codegreen},
 keywordstyle=\color{magenta},
    numberstyle=\tiny\color{codegray},
    stringstyle=\color{codepurple},
    basicstyle=\small\ttfamily,
    breakatwhitespace=false,         
    breaklines=false,                 
    captionpos=b,                    
    keepspaces=false,                 
    numbersep=5pt,                  
    showspaces=false,                
    showstringspaces=false,
    showtabs=false,                  
    tabsize=2,
    escapeinside=``
}

%% file: sec/intro.tex
\section{Introduction}
Automated Program Repair (APR) aims to automatically identify and correct bugs in source code, offering the potential to significantly reduce developer effort and improve software reliability~\shortcite{le2019automated}. Traditional APR techniques have predominantly focused on function-level APR, targeting small, isolated code fragments~\shortcite{gazzola2018automatic}. However, this narrow scope diverges from real-world software maintenance practices, which typically require reasoning over large codebases, interpreting complex issue descriptions, and addressing inter-file dependencies. Recent advances in Large Language Models (LLMs), particularly in code generation, have shifted attention toward applying LLMs to repository-level APR~\shortcite{bouzenia2024repairagent,zhang2024autocoderoverautonomousprogramimprovement, xia2024agentless, yang2024swe}. Unlike function-level APR, repository-level APR involves navigating large codebases, understanding intricate program logic, and identifying subtle bugs embedded across multiple files, introducing new and underexplored challenges.

To address these challenges, recent LLM-based APR methods fall into two main categories: agentic and procedural. Agentic approaches \shortcite{zhang2024autocoderoverautonomousprogramimprovement,yang2024swe,silva2023repairllama, DBLP:journals/corr/abs-2503-14269} treat the LLM as an autonomous agent that interacts with tools, navigates repositories, modifies code, and runs tests to resolve bugs. Procedural approaches \shortcite{xia2024agentless,ma2024alibaba}, in contrast, follow predefined pipelines, where the LLM sequentially performs tasks such as bug localization and patch generation using fixed prompts, without autonomous decision-making. While both paradigms show promise, they exhibit two key limitations.

\textbf{Neglecting Historical Repair Experience.} Current approaches treat each issue in isolation and primarily rely on feedback from the ongoing resolution process, while largely ignoring valuable historical experience accumulated from previously resolved issues within the same repository. In practice, software projects often exhibit recurring bug patterns~\shortcite{livshits2005dynamine}, such as incorrect API usage, type mismatches, or improper resource management. These patterns may manifest as syntactically similar changes or as diverse edits with similar underlying repair logic. Effectively capturing and leveraging historical repair knowledge could improve both repair efficiency and success rates.
    
\textbf{Limited Dynamic Adaptability Due to Static Prompting.} Current LLM-based repair methods rely on static, manually crafted prompts. However, these prompts fail to account for substantial variations across real-world software projects, {including differences in} coding conventions, architectural patterns, and project-specific practices. Manually customizing prompts per repository or issue type is impractical, particularly in large software ecosystems~\shortcite{DBLP:conf/iclr/HuCDZLYHTC25}. Moreover, static prompts cannot evolve in response to task feedback or adapt to nuanced context shifts during multi-step repair workflows. Addressing this limitation requires dynamic prompt generation or adaptation that tailors instructions based on repository characteristics, issue context, and insights from past repairs, enabling more robust, context-aware repair strategies.

In real-world software development, human programmers continuously accumulate knowledge from past debugging and repair activities, refining their strategies and reasoning over time. Inspired by this human-like learning process \shortcite{DBLP:conf/aaai/Zhao0XLLH24, DBLP:journals/corr/abs-2411-13941}, we explicitly incorporate historical repair experiences into repository-level APR, aiming to model how developers learn, adapt, and improve across issues within a software project. To effectively capture and reuse these accumulated experiences, we draw further inspiration from cognitive science. The dual-memory system theory~\shortcite{tulving1972episodic, tulving1985how} posits that human reasoning and decision-making rely on two complementary types of memory: episodic memory, which stores concrete personal experiences, and semantic memory, which captures abstract, generalizable knowledge. These two systems work synergistically, enabling individuals to recall specific events while applying broader insights to novel situations.

To this end, we propose \textsc{ExpeRepair}, a novel LLM-based APR approach that continuously accumulates and reuses historical repair experiences through a dual-memory system. {\tool} comprises two core modules: a Program Repair Module and a Memory Module. The repair module orchestrates a test agent and a patch agent, which collaboratively perform test generation, patch generation, and patch validation for each issue. The memory module organizes repair experiences into two complementary memories: episodic memory, which stores concrete repair demonstrations, and semantic memory, which distills abstract, natural language insights from past repair trajectories. At inference time, \textsc{ExpeRepair} retrieves relevant repair demonstrations and reflective insights to generate dynamic, context-aware prompts tailored to the current issue and repository. By maintaining both episodic memory and semantic memory, {\tool} achieves synergistic benefits: specific demonstrations provide precise implementation references, while reflective insights enable generalized adaptation to new and evolving contexts.

We evaluate our approach on SWE-Bench Lite and Verified~\shortcite{jimenez2023swe}, two challenging benchmarks consisting of real-world software issues sourced from popular GitHub repositories. Experimental results show that our approach outperforms all state-of-the-art open-source baselines, achieving pass@1 scores of 60.3\% and 74.6\% on SWE-Bench Lite and Verified, respectively, using Claude 4 Sonnet~\shortcite{claude7}. Furthermore, ablation studies validate the effectiveness of our dual-memory mechanism in enhancing repair performance. Our main contributions are as follows:

\begin{itemize}
    \item We propose \textsc{ExpeRepair}, an LLM-based APR method that continuously learns from historical repair experiences through dual-channel knowledge accumulation.
    \item We design a dynamic, experience-driven prompt generation mechanism that tailors repair strategies to the specific characteristics of each issue and repository. This mechanism addresses the limitations of static, handcrafted prompts by leveraging both concrete repair demonstrations and abstract repair patterns.
    \item We conduct comprehensive experiments on SWE-Bench, showing state-of-the-art performance and validating the contributions of each component through ablation studies.
    \item We provide publicly accessible source code \cite{website} to facilitate replication of our study and its application in broader contexts.
\end{itemize} 

%% file: sec/method.tex
\section{Method}
\label{sec:method}

\begin{figure*}[th]
\centering
\includegraphics[width=\textwidth]{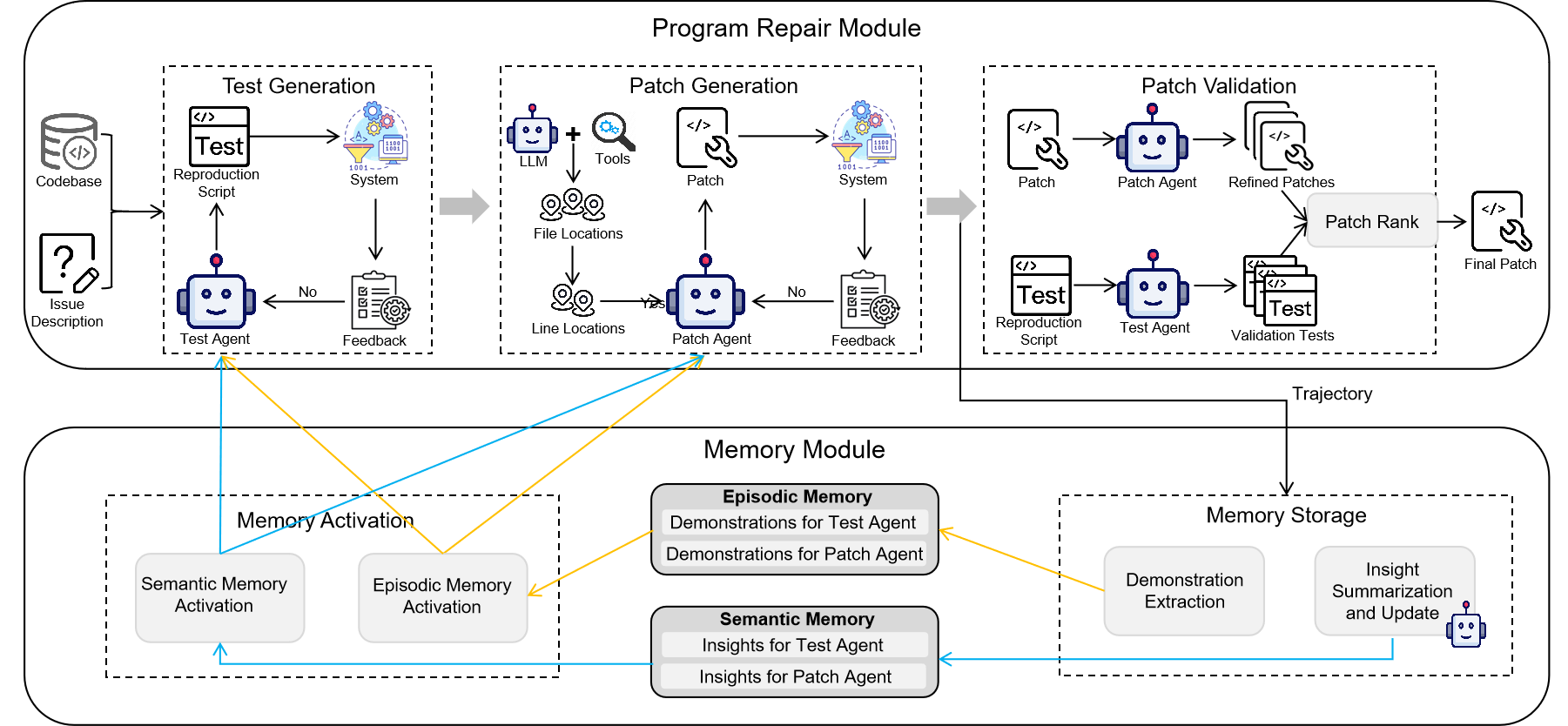} 
\caption{Overview of {\tool}.}
\label{fig:overview}
\end{figure*}

As illustrated in Figure~\ref{fig:overview}, {\tool} is built upon two key modules: the Program Repair Module and the Memory Module. The program repair module comprises a test agent and a patch agent, which collaboratively perform three essential tasks: test generation, patch generation, and patch validation. Once the program repair module processes an issue, the memory module captures the corresponding repair trajectory, extracting concrete demonstrations and summarizing high-level repair insights. These are stored in episodic memory and semantic memory, respectively. In subsequent repairs, {\tool} dynamically retrieves relevant demonstrations and insights to inform and enhance its repair strategies for new issues. In this section, we first present the end-to-end workflow of {\tool}, followed by a detailed illustration of its two core modules. 

\subsection{Workflow of {\tool}}
Conventional learning approaches~\shortcite{DBLP:journals/corr/abs-2308-10462, guo2024large, DBLP:journals/corr/abs-2308-12950, DBLP:journals/corr/abs-2401-14196} typically rely on computationally expensive fine-tuning to improve the capabilities of LLM-based agents. In contrast, {\tool} equips LLM agents with the dual-memory mechanism, enabling them to continuously acquire knowledge from their own repair experiences without requiring parameter updates. However, directly deploying {\tool} introduces a cold-start problem, where no historical repair experience is initially available to populate the memory systems. To overcome this, {\tool} adopts a two-phase strategy. In the initial phase, it is applied to a seed set of software issues to generate corresponding repair trajectories. The Memory Module then organizes these trajectories into two complementary memories. During this phase, {\tool} operates without accessing prior memories, and its prompts remain static. In the inference phase, {\tool} activates both memory systems: retrieving relevant demonstrations from episodic memory and recalling reflective insights from semantic memory. This enables {\tool} to dynamically compose context-aware, experience-driven prompts, progressively enhancing its repair capabilities.

\subsubsection{Initial Phase}
In this phase, {\tool} is applied to a set of seed issues. To ensure a fair comparison with baseline methods, which are evaluated directly on the SWE-Bench benchmark, we do not incorporate any external data as seed issues. Instead, {\tool} follows a straightforward approach: it attempts to reproduce the first five issues from each repository. If an issue is successfully reproduced (as determined by the LLM), {\tool} proceeds to perform a complete resolution. The resulting repair trajectories are then stored and organized within the memory module for subsequent recall. Specifically, the LLM agents in the program repair module use the ReAct algorithm~\shortcite{yao2023react} to iteratively process their respective tasks up to $Z$ times. Assuming an LLM agent processes the $n$-th issue at the $z$-th trial, it is provided with a task prompt $P$ defining its role and task (e.g., generating a patch), issue-specific context $C_n$ (e.g., the issue description or relevant code snippets), and feedback $F_{n, z-1}$ (e.g., execution results) from the $(z-1)$-th trial, which is initialized to the empty (i.e., $F_{n, 0} = \emptyset$). The agent then generates an output by sampling: $O_{n, z} \sim \mathrm{LLM}\left(P, C_n, F_{n, z-1}\right)$. After all agents complete their tasks, the memory module records their produced trajectories $\tau_{n, z} = [O_{n, 1}, F_{n, 1}, \dots, O_{n,z}, F_{n,z}]$, and subsequently extracts both issue-specific demonstrations and abstract natural language insights for each agent. These are stored in the episodic memory $\mathcal{M}_{\text{epi}}$ and the semantic memory $\mathcal{M}_{\text{sem}}$, which together serve as a springboard for a warm start.

\subsubsection{Inference Phase}
In this phase, {\tool} leverages accumulated experience to guide new issue repairs. Specifically, it first extracts summarized natural language insights $I$ from the semantic memory $\mathcal{M}_{\text{sem}}$, and appends them to the corresponding agent's task prompt $P$. It then forms a retrieval query $q_{n,z}=[C_n; F_{n,z-1}]$, which concatenates the current issue context with the latest feedback when available, and uses this query to retrieve the top-$k$ relevant demonstrations $D_{n, z} \subset \mathcal{M}_{\text{epi}}$. As a result, for the $n$-th issue at the $z$-th trial, each LLM agent is provided with the augmented input to generate the output: $O_{n, z} \sim \mathrm{LLM}\left(P + I, D_{n, z}, C_n, F_{n, z-1}\right)$. Similarly, after all agents complete their tasks, the memory module records the resulting repair trajectories, extracts new demonstrations and insights, and updates the dual memories. Over time, this process enables {\tool} to progressively enrich its episodic and semantic memories, fostering improvement in repair efficiency and effectiveness for future issues.

\subsection{Program Repair Module}

\subsubsection{Test Generation}
In practical software development, reproduction tests are essential for both demonstrating the presence of an issue and validating the correctness of candidate patches~\shortcite{qi2015analysis}. However, existing approaches typically prompt an LLM to generate a reproduction test script solely based on the issue description, which presents two major limitations: (1) frequent execution failures due to omitted dependencies, configurations, or environment-specific settings, and (2) inadequate issue reproduction, as such scripts often narrowly target the symptoms described in the issue without capturing the broader fault context.

{\tool} addresses these limitations by enabling the test agent to iteratively generate and refine reproduction tests based on dynamic feedback and memories accumulated from past repair trajectories. Specifically, before generating a new reproduction test at trial $z$ for the $n$-th issue, the test agent retrieves relevant demonstrations and extracts insights from both memories. Concretely, it forms the retrieval query $q^{\text{test}}_{n,z}=[C^{\text{test}}_n; F^{\text{test}}_{n,z-1}]$ and retrieves the top-$k$ demonstrations from episodic memory according to their similarity to this query:
$
D^{\text{test}}_{n,z} = \operatorname*{arg\,top\text{-}k}_{d \in \mathcal{M}_{\text{epi}}} \mathrm{Sim}\!\left(q^{\text{test}}_{n,z},\, q_d\right)
$
where $q_d$ is constructed from the issue context and feedback of each demonstration. This directly addresses the limitation (1): when the current test execution encounters a failure (e.g., a missing library or configuration error), demonstrations associated with the same or similar failures provide concrete examples of how those were successfully handled in the past. Then, the agent extracts all summarized natural language insights $I_{\text{test}}$ from the semantic memory $\mathcal{M}_{\text{sem}}$. These insights capture generalizable, high-level strategies distilled from prior repairs. {For example: ``When testing security-sensitive functionality, implement comprehensive test cases that verify proper handling of malicious inputs, edge cases, and potential attack vectors, ensuring robust validation and appropriate error responses'' (a real insight summarized by {\tool} for the SymPy project). These insights help address the limitation (2) by expanding the agent’s reasoning beyond narrowly reproducing stated symptoms.}

After recalling knowledge from the memories, {\tool} synthesizes the following adaptive prompt: (1) the enhanced task prompt $P^{\text{test}} + I^{\text{test}}$, (2) the issue description $C^{\text{test}}_n$, (3) the execution feedback $F^{\text{test}}_{n, z-1}$ from the previous test trial, and (4) the retrieved demonstrations $D^{\text{test}}_{n, z}$. Formally, the test agent's output at trial~$z$ is generated as: $O^{\text{test}}_{n, z} \sim \mathrm{LLM}\left(P^{\text{test}} + I^{\text{test}}, D^{\text{test}}_{n, z}, C^{\text{test}}_n, F^{\text{test}}_{n, z-1}\right)$. 
This iterative process continues until either a valid reproduction test is produced or $Z^{\text{test}}$ iterations have been completed.

\begin{promptbox}[Prompt for Test Generation]
Please analyze the issue description to understand the reported problem. Based on your analysis, write a standalone Python script \verb|`|test.py\verb|`| to reproduce the issue. \\
Requirements: \\
- The script should be minimal, including the essential tests necessary to demonstrate the issue. \\
- The script must be self-contained and runnable, including all necessary imports, dummy data, and setup. \\
- Use try-except blocks to isolate and execute each test case safely. \\
- Print the result of each test case (if applicable) to assist with debugging. \\

Here are reflective insights summarized from other issues in the same repository: \\
\texttt{<summarized\_insights>} \\

Below are example test scripts for other issues: \\
\texttt{<retrieved\_demonstrations>} \\

===\#\#\#==== \\
\texttt{<issue-specific context>} \\
\texttt{<feedback from the last trial>}
\end{promptbox}

\subsubsection{Patch Generation}
Following the successful reproduction of an issue, {\tool} initiates patch generation to produce candidate repairs that resolve the fault while preserving existing functionality. Before generating patches, it is essential to localize the issue, as accurate localization is a prerequisite for effective program repair. Following previous work \shortcite{yang2024swe}, {\tool} adopts a hierarchical localization strategy. Specifically, {\tool} first extracts the repository structure and provides it, together with the issue description, to the LLM to identify suspicious files. For each file, the LLM is subsequently prompted with the file content and the issue description to pinpoint likely faulty lines. The localization process is fully automated and relies on LLM-based reasoning rather than manually designed heuristics.

As noted in previous studies \shortcite{yang2024swe,bouzenia2024repairagent}, LLMs often generate incomplete patches, as fixing a repository-level issue typically requires coordinating changes across multiple parts of the code or performing a sequence of modifications. To address it, similar to the test generation process, {\tool} iteratively recalls episodic and semantic memories and leverages them to generate and refine patches. At each trial~$z$ for the $n$-th issue, the patch agent retrieves relevant demonstrations and extracts reflective insights, following the same procedure as in the test generation phase. The agent is then prompted with: (1) the enhanced task prompt $P^{\text{patch}} + I^{\text{patch}}$, (2) the issue description and localized faulty lines $C^{\text{patch}}_n$, (3) the latest execution feedback $F^{\text{patch}}_{n, z-1}$, and (4) the retrieved demonstrations $D^{\text{patch}}_{n, z}$. A candidate patch is generated as:
$
O^{\text{patch}}_{n, z} \sim \mathrm{LLM}\left(P^{\text{patch}} + I^{\text{patch}}, D^{\text{patch}}_{n, z}, C^{\text{patch}}_n, F^{\text{patch}}_{n, z-1}\right)
$. In each iteration, we perform multiple samplings with a high temperature to generate diverse candidate patches. This process continues until at least one candidate patch passes the reproduction test (i.e., the issue is no longer triggered) or reaches $Z^{\text{patch}}$ iterations.

\begin{promptbox}[Prompt for Patch Generation]
Your task is to analyze and resolve the given issue in two phases: \\
Phase 1: FIX ANALYSIS \\
1. Review the issue description and state clearly what the problem is. \\
2. Analyze the provided code context and specify where the problem occurs in the code. \\
3. State clearly the best practices to take into account in the fix. \\
4. State clearly how to fix the problem. \\

Phase 2: FIX IMPLEMENTATION \\
1. Focus on making minimal, precise, and relevant changes to resolve the issue. \\
2. Include any necessary imports introduced by the patch. \\
3. Write the patch using the strict format specified below: \\
\texttt{<patch\_format>} \\

Here are reflective insights summarized from other issues in the same repository: \\
\texttt{<summarized\_insights>} \\

Below are example patches for other issues: \\
\texttt{<retrieved\_demonstrations>} \\

===\#\#\#==== \\
\texttt{<issue-specific context>} \\
\texttt{<feedback from the last trial>}
\end{promptbox}

\subsubsection{Patch Validation}
After a candidate patch successfully passes the reproduction test, it is not immediately accepted as the final patch. This is because we empirically observed that reproduction scripts typically focus narrowly on the specific symptoms described in the issue, leading to false positives where patches pass limited tests but fail in broader contexts. To mitigate this, we revise and augment the patch by instructing the patch agent to address concerns such as edge case handling, regression risks, and adherence to language-specific best practices. Next, we augment the reproduction test suite by generating additional validation tests. The test agent is instructed to create tests targeting boundary conditions and corner cases. This process reduces the risk of narrow, brittle fixes that only address the reported symptom without addressing the underlying defect comprehensively.

To select the final patch, we first execute all candidate patches against the expanded test suite, which includes both reproduction and validation tests. Then, we pass candidate patches and their corresponding test results to the LLM, which selects the final patch based on criteria such as correctness and adherence to best practices.

\subsection{Memory Module}

\subsubsection{Memory Storage}
\label{method:store}
After the program repair module processes a new issue, the memory module collects the complete repair trajectory, which consists of all detailed artifacts associated with that specific repair instance, including the issue description, reproduction scripts, candidate patches, and the test execution results before and after each patch is applied. {{\tool} begins by extracting issue-specific demonstrations for both the test agent and the patch agent, which are stored in episodic memory. Specifically, for each trajectory, it collects two types of demonstrations: (1) \texttt{(input, successful output)} pairs, representing successful repair attempts, and (2) \texttt{(input, failed output, feedback, successful output)} tuples, capturing failed attempts along with their outputs, feedback, and corrected versions. The successful/failed output refers to whether the test script successfully reproduces the issue (as determined by the LLM) or whether the patch passes the reproduction test. All collected demonstrations are ultimately stored in episodic memory.}

{Then, {\tool} performs a summarize-and-update process to maintain the semantic memory, which stores high-level natural language insights. We prompt the LLM to analyze the complete repair trajectory and explain why specific attempts to reproduce the issue or generate a patch succeeded or failed. This generates reflective insights that may include general repair strategies, common failure causes, and effective heuristics. The newly generated insights are compared with those already stored in semantic memory, and {\tool} updates the memory by applying up to three operations: (1) \texttt{ADD}, to introduce a novel or valuable insight; (2) \texttt{REMOVE}, to discard contradictory, outdated, or redundant information; and (3) \texttt{EDIT}, to merge, generalize, or refine overlapping insights for improved clarity and utility. To maintain a manageable memory size, we impose a maximum number of insight entries per agent. When this limit is reached, {\tool} must first \texttt{REMOVE} an existing insight before applying a new \texttt{ADD}.}

\begin{promptbox}[Prompt for Insight Summarization and Update]
Your task is to:\\
1. Review the issue, analyze the insights provided by different Testers and Coders, and then summarize the new experiences for Testers and Coders, respectively.\\
2. Update the existing experiences by adding, editing, and removing them based on new summarized experiences.\\

To update the existing experiences, you must use the following operations:\\
\textbf{ADD:} Add a new experience if it is significantly different from existing experiences and could benefit future issues.\\
\textbf{REMOVE:} Remove an existing experience if it is CONTRADICTORY or SIMILAR/DUPLICATED to other existing experiences.\\
\textbf{EDIT:} Enhance or rewrite an existing experience to make it more general or valuable. If a new experience is similar to an existing one, merge them into a single improved experience using this operation.\\

IMPORTANT:\\
1. Align experiences strictly with the roles:\\
 - For Testers, include only experiences related to writing test cases.\\
 - For Coders, include only experiences related to developing patches, excluding any testing activities.\\
2. Ensure the new experience is practical, actionable, and broadly applicable to similar issues in the future.\\
3. Do not remove an existing experience solely because it is irrelevant to the current issue; it may still be useful for future cases. Aim to maintain a diverse set of experiences to help colleagues address a wide range of issues.\\
4. Perform at most four operations for tester and coder experiences individually, and each existing experience can receive only one operation.\\
5. The total number of experiences for each role must NOT EXCEED 15.
\end{promptbox}

\subsubsection{Memory Activation}
{To activate episodic memory, {\tool} uses the same retrieval query construction described above, namely $q_{n,z}=[C_n; F_{n,z-1}]$, and retrieves the top-$k$ most relevant past demonstrations using retrieval methods such as vector-based similarity (e.g., embedding search~\shortcite{DBLP:journals/pami/HjaltasonS03}) or term-based matching (e.g., BM25 \shortcite{DBLP:journals/ftir/RobertsonZ09}). The retrieved demonstrations are then incorporated into the agent's prompts as in-context examples, offering concrete precedents for constructing accurate reproduction scripts or patches. This memory mechanism enables {\tool} to avoid redundant exploration and benefit from accumulated experience, allowing it to generate higher-quality reproduction scripts and patches.}

Simultaneously, {\tool} activates the semantic memory by extracting all stored insights. The insights are incorporated into the task prompts for both agents to adaptively adjust and enhance their strategies in response to the characteristics of the current issue. This mechanism, effectively serving as a lightweight auto-prompting technique \shortcite{DBLP:conf/emnlp/WuGZZSYLDY24, DBLP:journals/corr/abs-2402-07927, DBLP:conf/emnlp/ShinRLWS20}, incrementally improves repair effectiveness by guiding agents with accumulated semantic experience.

%% file: sec/exp.tex
\section{Experimental Setup}

\subsection{Research Questions}
We address the following four research questions to assess the performance of {\tool}. 

\textbf{RQ1:} How effective is {\tool} in issue resolution compared to other approaches? 

\textbf{RQ2:} How does each component of {\tool} contribute to its overall performance? 

\textbf{RQ3:} How does the choice of underlying LLM affect {\tool}'s performance?

\textbf{RQ4:} How sensitive is {\tool} to key hyperparameters in its memory module?

\subsection{Benchmark}
To evaluate the effectiveness of {\tool} in addressing real-world software issues, we conduct experiments on two subsets of the SWE-Bench benchmark~\cite{jimenez2023swe}: {SWE-Bench Lite} and {SWE-Bench Verified}. {SWE-Bench Lite} consists of 300 GitHub issues sampled from real-world Python projects. SWE-Bench Verified~\cite{sweverified}, produced by OpenAI, is a curated subset in which each issue has been validated by human developers to ensure it contains sufficient information to be solvable.

\subsection{Baselines}
Repository-level program repair has received significant attention from both academia and industry. Numerous approaches have been proposed and evaluated on SWE-Bench. In this work, we focus on open-source methods, as the technical details of closed-source systems are generally unavailable. We compare our approach against the following representative open-source baselines:

\begin{itemize}
    \item \textbf{SWE-agent} \cite{yang2024swe} introduces a general agent–computer interface that enables LLMs to autonomously leverage external tools for fixing issues in real GitHub repositories.
    
    \item \textbf{Aider} \cite{Aider} is an LLM-based coding assistant that combines repository-aware context construction with Git-integrated editing to produce context-aware fixes in large codebases.
    
    \item \textbf{AutoCodeRover} \cite{zhang2024autocoderoverautonomousprogramimprovement} is an autonomous program improvement system that merges LLMs with advanced code search capabilities. It addresses GitHub issues through iterative patch generation and refinement.
    
    \item \textbf{SpecRover} \cite{DBLP:conf/icse/Ruan0R25} extends AutoCodeRover by incorporating code intent extraction via LLMs, enhancing issue resolution. On the full SWE-Bench, SpecRover achieves over 50\% improvement in efficacy compared to AutoCodeRover, demonstrating its advanced autonomous repair capabilities.
    
    \item \textbf{Agentless} \cite{xia2024agentless} follows a procedural approach that decomposes the repair process into distinct phases: localization, repair, and patch validation.
    
    \item \textbf{OpenHands} \cite{wang2024openhands} is a comprehensive framework that leverages LLMs for real-world software engineering tasks. It integrates a sandboxed execution environment, a critic model for ranking solutions, and a dual-agent architecture with experience banks to improve repair performance.
    
    \item \textbf{PatchPilot} \cite{li2025patchpilot} is a procedural repair method that balances efficacy, stability, and cost-efficiency. It guides bug repair through a five-step workflow: reproducing the error, localizing the fault, generating a fix, validating the result, and refining the solution.
    
    \item \textbf{DARS} \cite{DARS} introduces an inference-time compute scaling strategy for coding agents. By branching only at critical decisions and providing long-horizon feedback, it reduces resets and accelerates execution, improving overall efficiency.
\end{itemize}

\subsection{Metrics}
Following prior work~\cite{yang2024swe, zhang2024autocoderoverautonomousprogramimprovement, xia2024agentless}, we evaluate the performance of the baselines and {\tool} using the following primary metrics:

\begin{itemize}
    \item \textbf{pass@1:} The percentage of issues successfully resolved on the first attempt. This metric reflects the method's ability to generate correct patches without requiring multiple submissions. It represents the most practical scenario for real-world deployment.
    
    \item \textbf{Average Cost:} The average inference cost incurred when executing the repair method. This metric captures the computational efficiency and resource requirements of each approach, providing a practical measure of its operational overhead.
\end{itemize}

\noindent
These primary metrics form the basis for the overall performance comparison in Section~\ref{main_result}. In addition, we introduce two supplementary metrics in our ablation studies to provide deeper insights into the intermediate stages of the repair process and to better explain the performance differences:

\begin{itemize}
    \item \textbf{Execution Success Rate (ESR):} The percentage of generated reproduction scripts that execute successfully without errors, such as missing dependencies or configuration issues.
    
    \item \textbf{Reproduction Success Rate (RSR):} The percentage of reproduction scripts that correctly reproduce the target issue, as manually verified by human annotators.
\end{itemize}

\subsection{Implementation}
\label{sec:implementation}
{\tool} uses Claude~3.5/4~Sonnet as its primary foundation model throughout the pipeline. o4-mini is invoked only in rare fallback cases (less than 1\% of total LLM calls) when temporary API unavailability occurs, following a practice similar to SpecRover \cite{DBLP:conf/icse/Ruan0R25}. Due to space limitations, we present only the key prompts for test generation, patch generation, and memory storage in Section~\ref{sec:method}. The remaining prompts are available on our project website~\cite{website}. Below, we summarize the implementation details for each phase.

\textbf{Test and Patch Generation.}  
For each issue, the test agent iteratively generates a reproduction script for up to 3 rounds. Upon successfully reproducing the issue, the patch agent performs up to 3 iterations of patch generation, producing 4 candidate patches per iteration. Failed patches and execution results are fed back to guide new candidates, while passing patches proceed to validation.

\textbf{Patch Validation.}  
To reduce false positives, {\tool} samples 3 validation tests by prompting the LLM to generate edge cases and boundary conditions from the reproduction script. The patch agent then generates 4 augmented patches addressing these cases, regression risks, and language-specific best practices. All candidates are executed against validation tests, and an LLM reviews the results to select the best-performing patch.

\textbf{Memory Storage.}  
For episodic memory, demonstrations are stored along with metadata (issue description, status, timestamp), as described in Section~\ref{method:store}. Semantic memory stores key insights summarized by the LLM. These insights are compared against existing entries and updated via \texttt{ADD}, \texttt{REMOVE}, or \texttt{EDIT}. The number of stored insights is capped at 15. When this limit is reached, the LLM removes insights deemed outdated or less reliable to make room for new ones.

\textbf{Memory Activation.}  
For episodic memory, {\tool} retrieves the top 3 most relevant demonstrations via the BM25 algorithm, which are then used as in-context examples for the agents. For semantic memory, stored insights are appended to the prompts as high-level repair guidelines, enhancing the agents' reasoning and decision-making capabilities.

%% file: sec/results.tex
\section{Results}
\label{sec:result}

\subsection{Overall Performance}
\label{main_result}
We conduct a comparative evaluation of {\tool} against a set of recent open-source baselines on the SWE-Bench Lite and SWE-Bench Verified benchmarks. Note that the first five issues, which are used to initialize the semantic and episodic memory components, are also included in the total issues addressed in the standard sequential order of the evaluation to compute the final resolution rate. This design ensures repair performance is evaluated on all issues and provides a fair comparison with all baselines. Table~\ref{tab:rq1} reports the resolution rate (pass@1) and the average inference cost per instance (USD) for all methods.

Overall, {\tool} achieves state-of-the-art performance across both benchmarks. Using Claude 3.5 Sonnet, {\tool} resolves 48.3\% of issues on SWE-Bench Lite and 57.2\% on SWE-Bench Verified, surpassing all previously published open-source approaches, including PatchPilot~\cite{li2025patchpilot} and DARS~\cite{DARS}. When paired with Claude 4 Sonnet, {\tool} further improves performance, achieving 60.3\% resolution on SWE-Bench Lite and 74.6\% on SWE-Bench Verified. These results demonstrate both the adaptability of {\tool} and the potential benefits of leveraging more powerful underlying LLMs.

In addition to effectiveness, {\tool} maintains a competitive inference cost. For example, while DARS achieves a pass@1 of 47.0\% on SWE-Bench Lite, its average cost is 12.24 USD per instance, which is roughly six times higher than {\tool}'s 1.89 USD using Claude 3.5 Sonnet. This highlights that {\tool} not only improves repair effectiveness but also offers practical efficiency suitable for real-world deployment. Compared with other strong baselines, {\tool} consistently outperforms PatchPilot and OpenHands, which achieve 45.3\% and 41.7\% resolution on SWE-Bench Lite, respectively, with comparable or slightly higher costs. Similarly, Agentless-1.5 attains 40.7\% resolution with an inference cost of 1.12 USD, indicating that while it is cost-efficient, its effectiveness remains substantially lower than that of {\tool}.

\input{tab/rq1}

\mybox{
\textbf{Answering RQ1:} {\tool} achieves state-of-the-art performance on both SWE-Bench Lite and SWE-Bench Verified, resolving 48.3\% and 57.2\% of issues with Claude 3.5 Sonnet, and up to 60.3\% and 74.6\% with Claude 4 Sonnet. It consistently outperforms all open-source baselines while maintaining competitive inference costs, demonstrating a favorable balance between effectiveness and efficiency.
}

\subsection{Ablation Study}
To assess the contribution of individual components in {\tool}, we conduct a systematic ablation study on the SWE-Bench Lite and SWE-Bench Verified benchmarks. Unless otherwise specified, all experiments use Claude~3.5 Sonnet as the foundation model.

\subsubsection{Ablations on Episodic Memory and Semantic Memory Components.}
We evaluate the contributions of Episodic and Semantic Memory by disabling them individually or jointly:  
(1) \textbf{w/o Memory Module}, which removes both Episodic and Semantic Memory and treats each issue in isolation, as
in prior methods;
(2) \textbf{w/o Episodic Memory}, which removes the in-context demonstrations provided to LLM agents during test and patch generation; and
(3) \textbf{w/o Semantic Memory}, which removes reflective insights distilled from prior repair experiences.

\input{tab/major_ablation}

As shown in Table~\ref{tab:ablation_study}, removing the Memory Module leads to the most significant performance degradation, with the resolution rate dropping from 48.3\% to 42.3\% on SWE-Bench Lite and from 57.2\% to 50.4\% on SWE-Bench Verified. The Execution Success Rate (ESR) and Reproduction Success Rate (RSR) also decrease substantially, by 20.7 and 21.0 pp on Lite, and by 25.0 and 25.2 pp on Verified. This underscores the importance of leveraging past repair knowledge to guide LLM agents in reproducing issues and generating correct patches.

Disabling Episodic Memory also degrades performance notably. The pass@1 decreases to 44.7\% on SWE-Bench Lite and 52.8\% on SWE-Bench Verified, accompanied by ESR and RSR reductions of 14.0/13.7 pp on Lite and 16.2/17.6 pp on Verified. This suggests that the demonstrations stored in the episodic memory provide effective in-context guidance during both test and patch generation. In contrast, removing the Semantic Memory leads to a relatively smaller performance drop. The resolution rates decrease modestly to 47.0\% on Lite and 56.2\% on Verified, while ESR and RSR drop by 4.7/6.4 pp on Lite and 6.4/5.6 pp on Verified. This indicates that while the insights distilled in semantic memory improve the quality of generated patches, their effect is less pronounced compared to issue-specific demonstrations.

Overall, these results reveal a clear hierarchy in component contributions: the Memory Module has the largest impact, followed by Episodic Memory, and then Semantic Memory. The combination of Episodic and Semantic Memory yields a synergistic effect, producing the highest pass@1, ESR, and RSR metrics across both benchmarks. These findings validate the design rationale of {\tool}: leveraging prior experience to provide both in-context demonstrations and reflective insights delivers substantial gains in autonomous program repair.

\subsubsection{Ablations on Episodic Memory Retrieval.}

\begin{figure*}[th]
\centering
\includegraphics[width=0.85\textwidth]{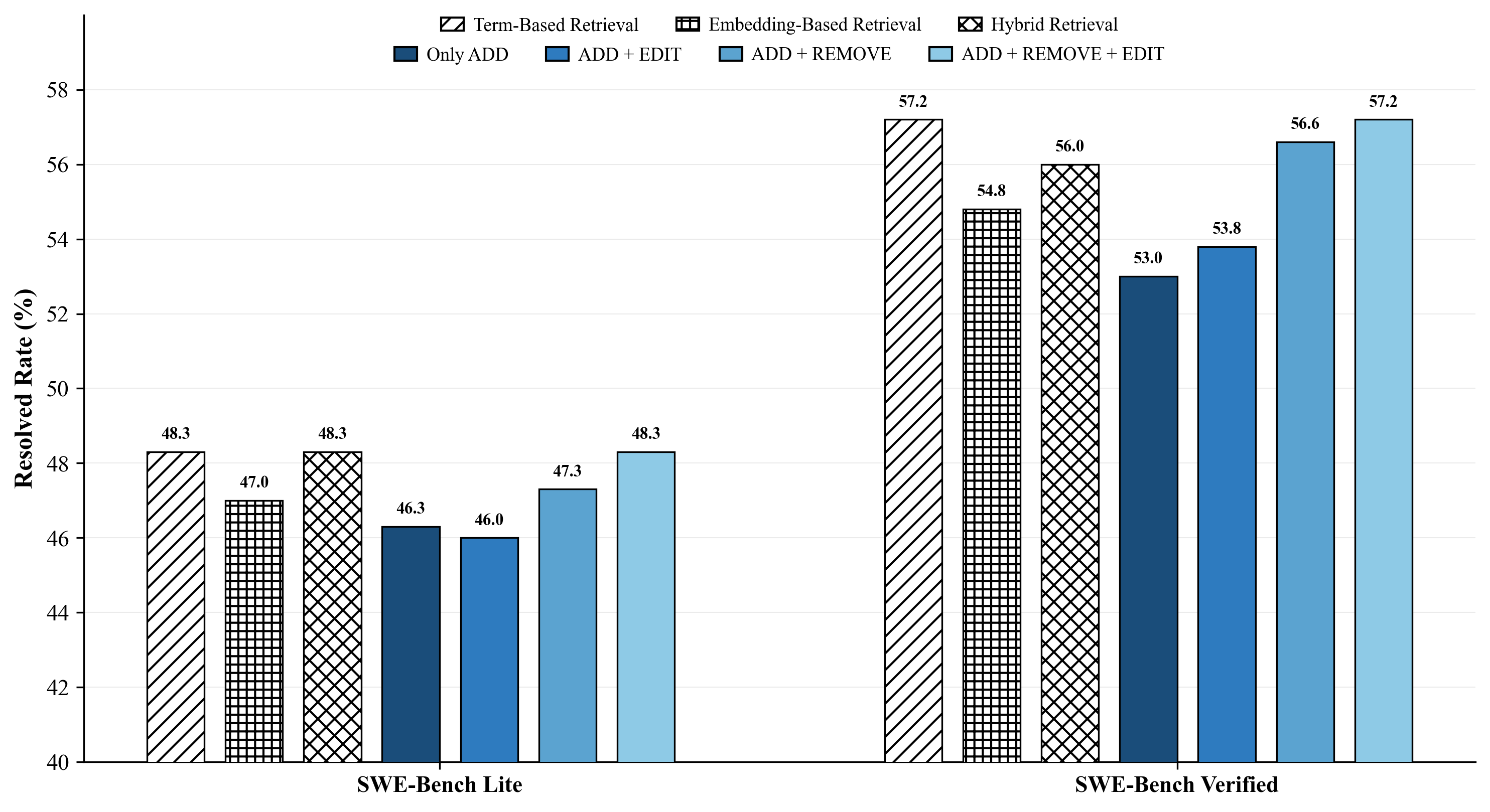} 
\caption{{Ablations on memory retrieval and update strategy}.}
\label{fig:ablation_major}
\end{figure*}

We evaluate alternative retrieval strategies for Episodic Memory. In addition to the \textbf{Term-Based Retrieval} method (BM25) used in {\tool}, we evaluate an \textbf{Embedding-Based Retrieval} approach that computes cosine similarity between embeddings generated by the Multilingual-E5-Large~\cite{wang2024multilingual} model. We also consider a \textbf{Hybrid Retrieval} strategy that linearly combines BM25 and embedding similarity scores, weighted by $\alpha$ and $1-\alpha$, respectively. We sweep $\alpha \in \{0.3, 0.4, 0.5, 0.6, 0.7\}$ and report the best-performing results.

Figure~\ref{fig:ablation_major} shows that BM25 achieves the best or tied-best pass@1 on both SWE-Bench Lite (48.3\%) and SWE-Bench Verified (57.2\%). Embedding-based retrieval underperforms, while the hybrid approach remains competitive but does not surpass BM25. We attribute this to the importance of lexical overlap in failure signals (e.g., error messages, stack traces, and test feedback), which are better captured by term-based matching. Dense retrieval occasionally returns semantically related yet causally different issues. These findings align with prior work~\cite{DBLP:journals/tosem/YangCGLHLX25}, suggesting that more complex retrieval does not necessarily improve code generation.

\subsubsection{Ablations on Semantic Memory Update Mechanisms.}

To further investigate the impact of different Semantic Memory maintenance strategies, we evaluate four variants:
(1) \textbf{Only ADD}, where new insights are appended without modification or deletion;
(2) \textbf{ADD+EDIT}, which allows existing entries to be revised but not removed;
(3) \textbf{ADD+REMOVE}, which supports insertion and deletion without editing; and
(4) \textbf{ADD+EDIT+REMOVE}, the full strategy used in {\tool}. For all variants, when the accumulated semantic insights exceed the model's input limit, we truncate the oldest entries as needed to stay within the input budget.

Figure~\ref{fig:ablation_major} shows that variants without \texttt{REMOVE} consistently underperform, as accumulated insights can saturate the prompt context and introduce noise. Incorporating \texttt{REMOVE} yields substantial gains, particularly on SWE-Bench Verified, improving pass@1 by 3.6 pp over \texttt{ADD}-only. This underscores the importance of pruning obsolete or misleading entries to maintain memory quality.

While \texttt{EDIT} alone provides limited improvement, combining it with \texttt{REMOVE} achieves the best performance across both benchmarks. This suggests a complementary effect: \texttt{REMOVE} acts as the primary mechanism for controlling memory growth and eliminating noise, \texttt{EDIT} further improves memory quality by refining the remaining entries.

\mybox{
\textbf{Answering RQ2:} The ablations confirm that {\tool}'s gains stem from its dual-memory design. Removing the Memory Module substantially degrades pass@1 on both benchmarks. Episodic Memory contributes the majority of improvements through issue-specific repair demonstrations, while Semantic Memory provides consistent complementary gains via accumulated insights. Further analysis shows that effective memory design requires both accurate retrieval of relevant past experiences and disciplined long-term memory maintenance, with term-based retrieval and incremental memory updates yielding the best overall performance.
}


\subsection{Results on Different LLMs}

\begin{wrapfigure}{r}{0.50\textwidth}
    \centering
    \includegraphics[width=0.48\textwidth]{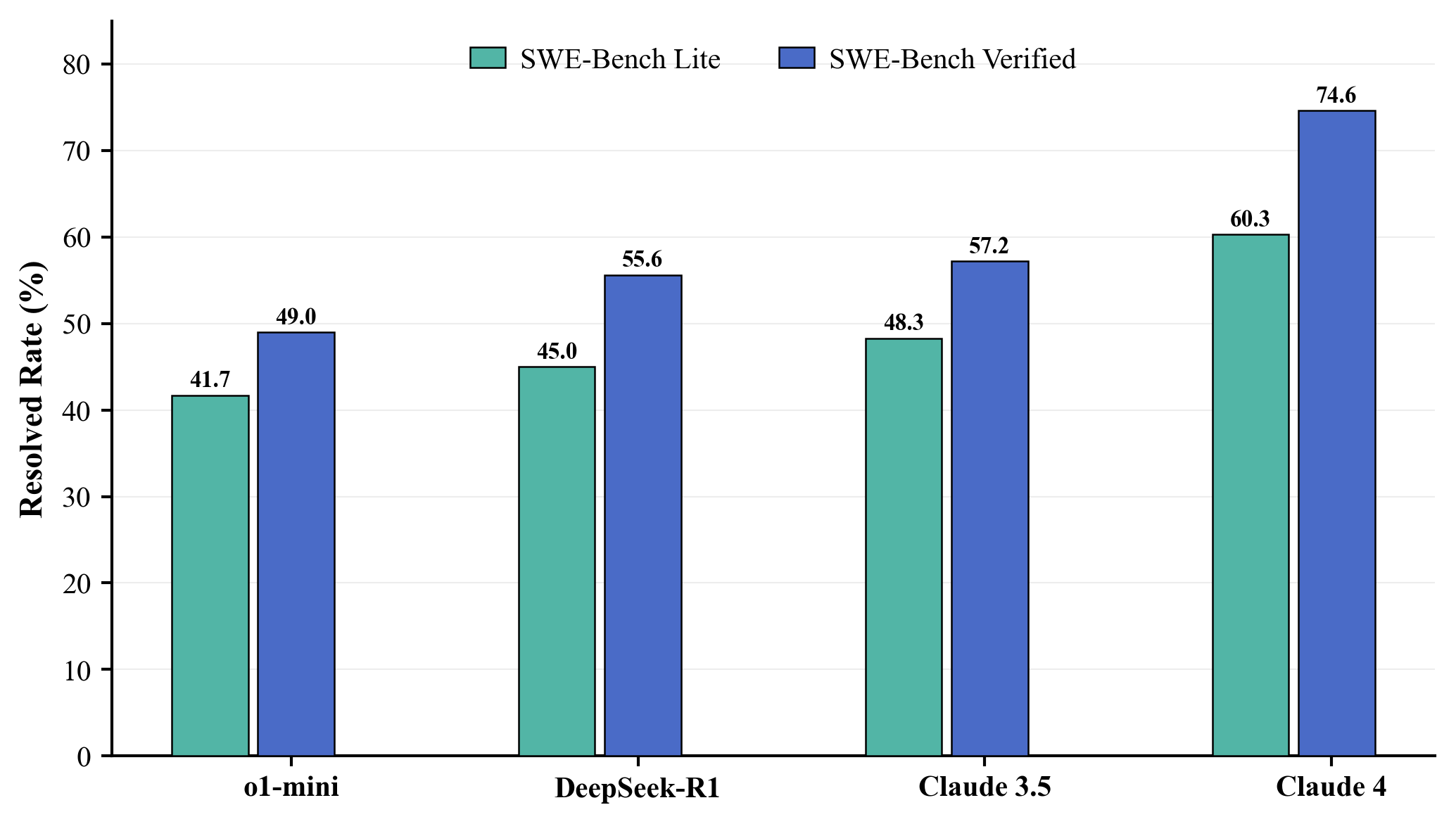}
    \caption{{Results of {\tool} on different models}.}
    \label{fig:different_models}
\end{wrapfigure}

To evaluate {\tool}'s performance across different foundation models, we integrate it with four representative models: o1-mini~\shortcite{o1mini}, DeepSeek-R1~\shortcite{deepseek}, Claude 3.5 Sonnet~\shortcite{claude5}, and Claude 4 Sonnet~\shortcite{claude7}, while keeping all other components unchanged. As noted in Section~\ref{sec:implementation}, o4-mini is invoked only when a query encounters a temporary API error. For brevity, the x-axis labels in Figure~\ref{fig:different_models} list only the different foundation models.

Figure~\ref{fig:different_models} presents the results. Across both benchmarks, we observe a consistent performance ranking among the evaluated models. Claude 4 Sonnet achieves the best performance, followed by Claude 3.5 Sonnet, DeepSeek-R1, and o1-mini. This indicates that {\tool} preserves the relative strengths of different models and benefits from stronger underlying models.

pass@1 on SWE-Bench Lite ranges from 41.7\% (o1-mini) to 60.3\% (Claude 4 Sonnet), and from 49.0\% to 74.6\% on SWE-Bench Verified.  Notably, even when paired with relatively less capable models such as o1-mini and DeepSeek-R1, {\tool} still achieves competitive results. For example, {\tool} with o1-mini attains pass@1 performance comparable to SpecRover built on the Claude 3.5 Sonnet model.

\mybox{
\textbf{Answering RQ3:} {The experiments demonstrate that {\tool} is broadly compatible with diverse LLM backbones across both benchmarks. Furthermore, stronger base models consistently achieve higher performance, underscoring the importance of model selection when deploying automated code repair systems in practice.}
}

\subsection{Sensitivity Analysis of Hyperparameters}

{In this section, we analyze the sensitivity of {\tool} to two key hyperparameters: (1) the number of demonstrations retrieved from episodic memory, and (2) the number of insights stored in semantic memory. Each parameter is varied independently while keeping all other settings fixed. All experiments are conducted on SWE-Bench Lite and SWE-Bench Verified using Claude~3.5 Sonnet as the backbone model.}

\subsubsection{Number of Retrieved Demonstrations.}

\begin{figure*}[tbh]
\centering
\includegraphics[width=0.9\textwidth]{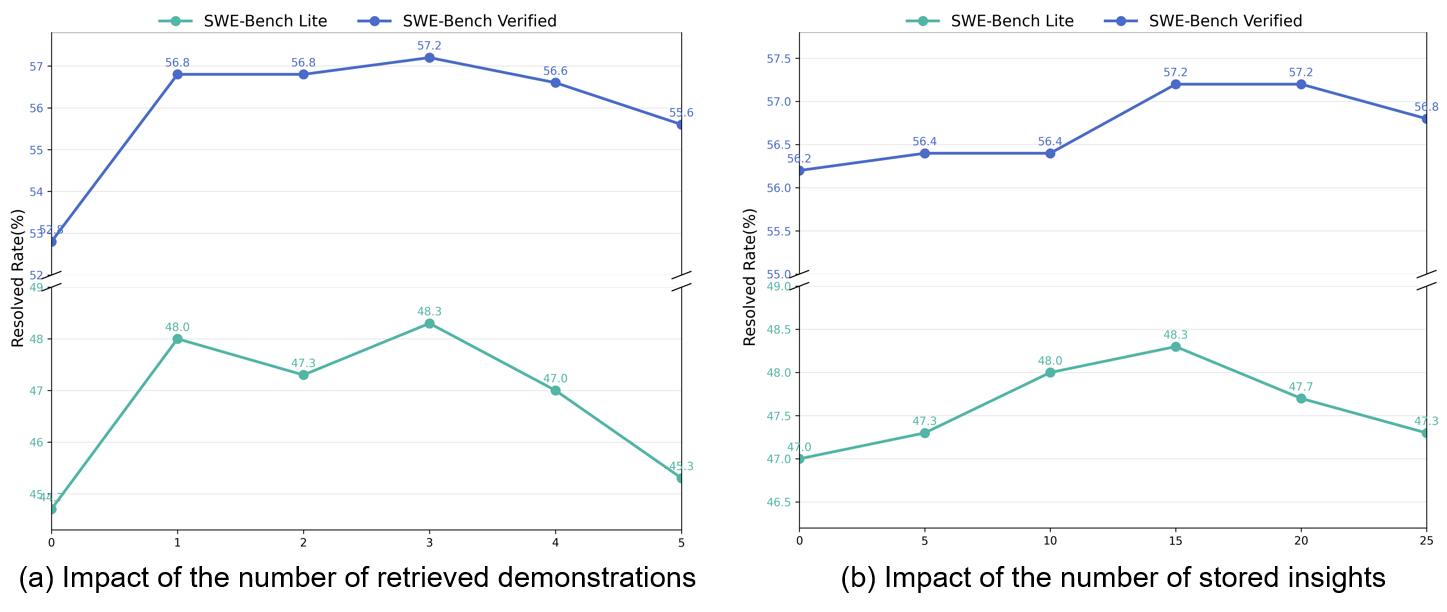}
\caption{{Sensitivity analysis of the number of retrieved demonstrations and stored insights.}}
\label{fig:sensitivity}
\end{figure*}

{\tool} retrieves the top-$k$ most relevant demonstrations to enhance the agent’s reasoning process. We vary $k$ from 0 (no demonstrations) to 5. The results are shown in Figure~\ref{fig:sensitivity}(a).

Across both benchmarks, introducing at least one retrieved demonstration leads to substantial improvements over $k=0$. Specifically, performance increases from 44.7\% to 48.0\% on Lite and from 52.8\% to 56.8\% on Verified when increasing $k$ from 0 to 1. This confirms the important role of episodic memory in improving program repair performance through leveraging prior task examples.

Performance peaks at $k=3$ on both datasets, reaching 48.3\% on Lite and 57.2\% on Verified. Further increasing $k$ degrades results (45.3\% and 55.6\% at $k=5$), suggesting a trade-off between retrieval breadth and contextual interference. While retrieving more demonstrations increases the likelihood of including relevant examples, it also raises the risk of incorporating less relevant content and consuming additional context budget, which may dilute attention to the current issue.

\subsubsection{Number of Stored Insights.}

{We vary the number of stored insights from 0 to 25 in increments of 5, as shown in Figure~\ref{fig:sensitivity}(b).}

{On SWE-Bench Lite, performance improves from 47.0\% (no insights) to a peak of 48.3\% at 15 insights, followed by a gradual decline to 47.3\% at 25 insights. A similar pattern is observed on Verified, where performance increases from 56.2\% to 57.2\% at 15–20 insights, before slightly decreasing to 56.8\% at 25 insights.}

These results indicate that semantic memory is beneficial up to a moderate size (approximately 15–20 insights), after which gains plateau or slightly diminish, likely due to redundancy or lower-utility information within the constrained context window.

\mybox{
{\textbf{Answering RQ4:} {\tool} demonstrates stable and robust behavior across a reasonable range of hyperparameter settings. The empirically best configuration, i.e., 15 stored insights and top-3 demonstration retrieval, achieves 48.3\% on Lite and 57.2\% on Verified. Importantly, performance remains near-optimal within moderate deviations (10–20 insights and $k=2$–4), indicating that the method is not overly sensitive to minor hyperparameter variations.}
}

%% file: tab/rq1.tex
\begin{table*}[t]
\renewcommand\arraystretch{1.2}
\centering
\caption{Comparison of {\tool} and open-source baselines on SWE-Bench Lite and Verified benchmarks. For baselines, we directly use the reported results either from the official leaderboard or from the tool's official paper/repository.}
\resizebox{0.9\textwidth}{!}{
\begin{tabular}{cc|cc|cc}
\toprule
\multirow{2}{*}{\textbf{Method}} & \multirow{2}{*}{\textbf{LLM}} & \multicolumn{2}{c|}{\textbf{SWE-Bench Lite}}                                   & \multicolumn{2}{c}{\textbf{SWE-Bench Verified}} \\ \cline{3-6} 
                                 &                     & \textbf{pass@1 (\%)} & \textbf{AVG Cost (\$)} & \textbf{pass@1 (\%)} & \textbf{AVG Cost (\$)}     \\ \hline 
SWE-agent                        & Claude 3.5 Sonnet                     & 23.0                   & {1.62}                   &  33.6                  &  1.59                 \\
Aider                    & GPT-4o + Claude 3 Opus                                  & 26.3                 &  -                  &  -                 & -                  \\
AutoCodeRover                 & GPT-4o                                   & 30.7                   & -                   & -                   & -                  \\
SpecRover                   & Claude 3.5 Sonnet + GPT-4o                                   & 31.0                   & 0.65                   & 46.2                 & -                \\
Agentless-1.5                    & Claude 3.5 Sonnet                     & 40.7                   & {1.12}                   &  50.8                  &  1.19                 \\
OpenHands                        & CodeAct v2.1                             & 41.7                   & 1.33                   & 53.0                   &  0.78                  \\
PatchPilot                       & Claude 3.5 Sonnet                    & 45.3                   & {0.97}                   &   53.6                  &   0.99                \\
DARS                             & Claude 3.5 Sonnet + DeepSeek R1                    & 47.0                   & {12.24}                  &  -                  &  -                 \\ \hline
\multirow{2}{*}{\tool}            & Claude 3.5 Sonnet + o4-mini                    & 48.3                   & 1.89                   &  57.2                  & 1.74                  \\
            & Claude 4 Sonnet + o4-mini                   & 60.3                   &   1.96                 &   74.6                 &  1.91                \\
\bottomrule
\end{tabular}
}
\label{tab:rq1}
\end{table*}

%% file: tab/major_ablation.tex
\begin{table*}[t!]
\renewcommand\arraystretch{1.2}
\centering
\caption{Ablations on Episodic Memory and Semantic Memory components}
\resizebox{0.9\textwidth}{!}{
\begin{tabular}{lcccccc}
\toprule
\multicolumn{1}{c}{\multirow{2}{*}{\textbf{Method}}}
& \multicolumn{3}{c}{\textbf{SWE-Bench Lite}} 
& \multicolumn{3}{c}{\textbf{SWE-Bench Verified}} \\
\cmidrule(l){2-7}
& \textbf{pass@1 (\%)} & \textbf{ESR (\%)} & \textbf{RSR (\%)} 
& \textbf{pass@1 (\%)} & \textbf{ESR (\%)} & \textbf{RSR (\%)} \\
\hline
\textbf{{\tool}}
& 48.3 & 82.0 & 78.7 
& 57.2 & 84.6 & 82.2 \\
\quad w/o Memory Module
& 42.3 & 61.3 & 57.7 
& 50.4 & 59.6 & 57.0 \\
\quad w/o Episodic Memory
& 44.7 & 68.0 & 65.0 
& 52.8 & 68.4 & 64.6 \\
\quad w/o Semantic Memory
& 47.0 & 77.3 & 72.3 
& 56.2 & 78.2 & 76.6 \\
\bottomrule
\end{tabular}
}
\label{tab:ablation_study}
\end{table*}

%% file: sec/discussion.tex
\section{Discussion}
\subsection{Case Study}
\begin{figure}[htbp]
\centering
\includegraphics[width=0.9\textwidth]{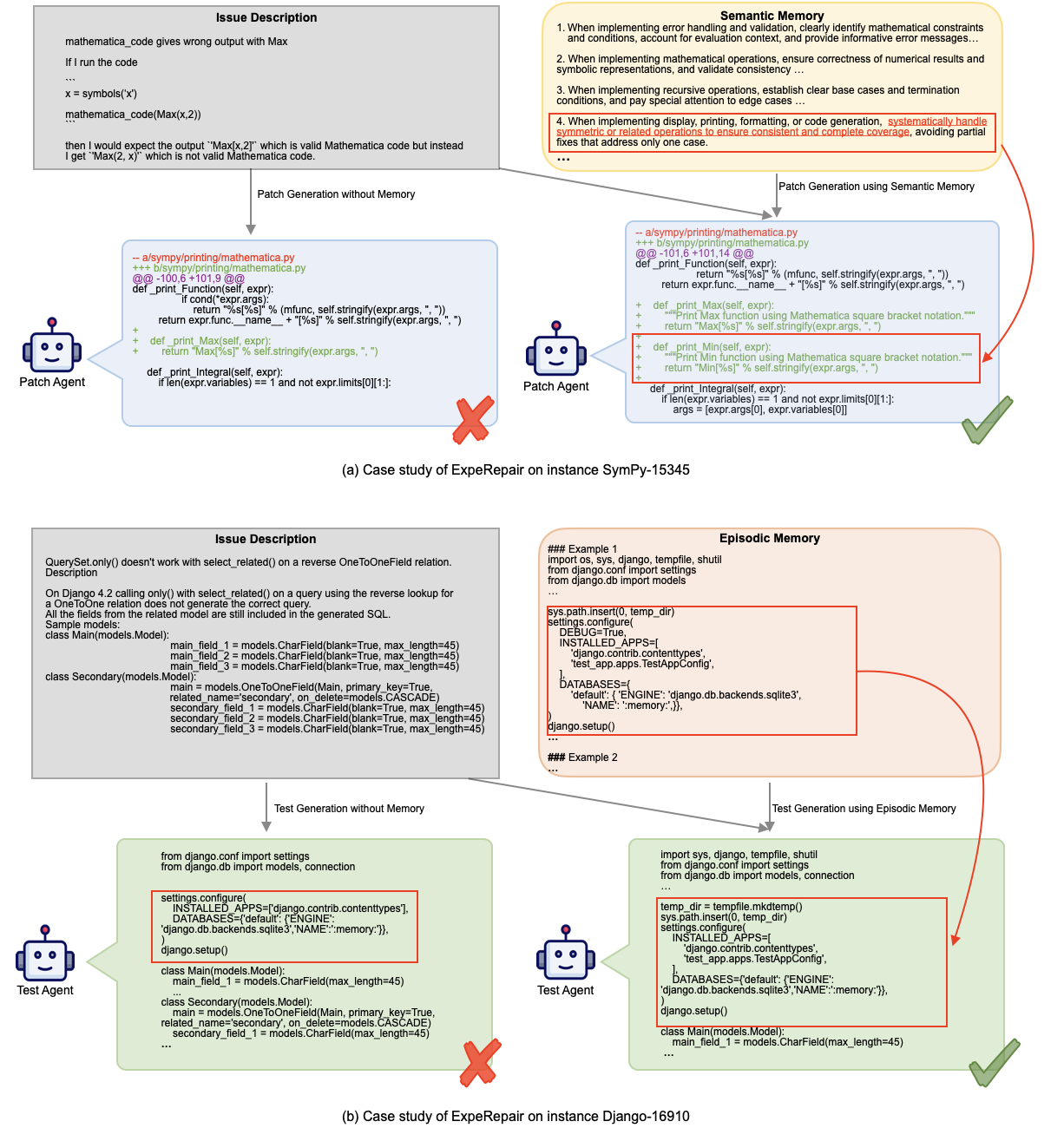}
\vspace{-0.2cm}
\caption{Two real cases from the SymPy and Django repositories resolved by {\tool}.}
\vspace{-0.3cm}
\label{fig:case_study}
\end{figure}

To further evaluate {\tool}'s effectiveness, we present two real-world cases showing how semantic and episodic memory improve agents' decisions in bug reproduction and patch generation.

\subsubsection{Semantic Memory}
Figure \ref{fig:case_study} (a) illustrates an issue in \texttt{SymPy} related to incorrect code generation for the \texttt{Max} function when using the \texttt{mathematica\_code} printer. Specifically, given the input
\[
\texttt{x = symbols(`x');\quad mathematica\_code(Max(x,2))}
\]
the expected output is \texttt{Max[x,2]}, which conforms to valid Mathematica syntax. Instead, the current implementation produces \texttt{Max(2, x)}, which violates Mathematica's function-call conventions. This behavior arises because the \texttt{mathematica\_code} printer mishandles variadic functions such as \texttt{Max}, generating syntactically invalid code. Although the issue is framed in terms of \texttt{Max}, the same underlying problem affects its counterpart function \texttt{Min}. This highlights the need for solutions that systematically handle symmetric or related operations to ensure consistent code generation.  

Initially, the patch agent attempted to resolve the issue without activating semantic memory. The generated patch introduced a specialized \texttt{\_print\_Max} method to enforce Mathematica's square-bracket notation. While this patch correctly formats \texttt{Max}, it fails to address the counterpart function \texttt{Min}, leaving it incorrectly printed. Consequently, although \texttt{Max} expressions are valid, the patch produces asymmetric and incomplete code generation.  

Subsequently, the patch agent was allowed to activate semantic memory, incorporating all previously stored reflective insights into its system prompt. Notably, the fourth insight, 
\textit{"When implementing display, printing, formatting, or code generation, systematically handle symmetric or related operations to ensure consistent and complete coverage, avoiding partial fixes that address only one case."}, 
which was derived from resolving similar issues such as SymPy-13031, guides the agent to consider not only the explicitly reported \texttt{Max} function but also \texttt{Min} and other symmetric operators. As a result, the newly generated patch resolves the reported \texttt{Max} defect and simultaneously addresses \texttt{Min}, yielding a robust and comprehensive solution. This ensures consistent and syntactically valid Mathematica code generation across all relevant variadic functions, effectively preventing asymmetric or incomplete code generation.

\subsubsection{Episodic Memory}
\begin{figure}[thbp]
    \centering
    \begin{minipage}[b]{0.48\textwidth}
        \centering
        \includegraphics[height=5.5cm]{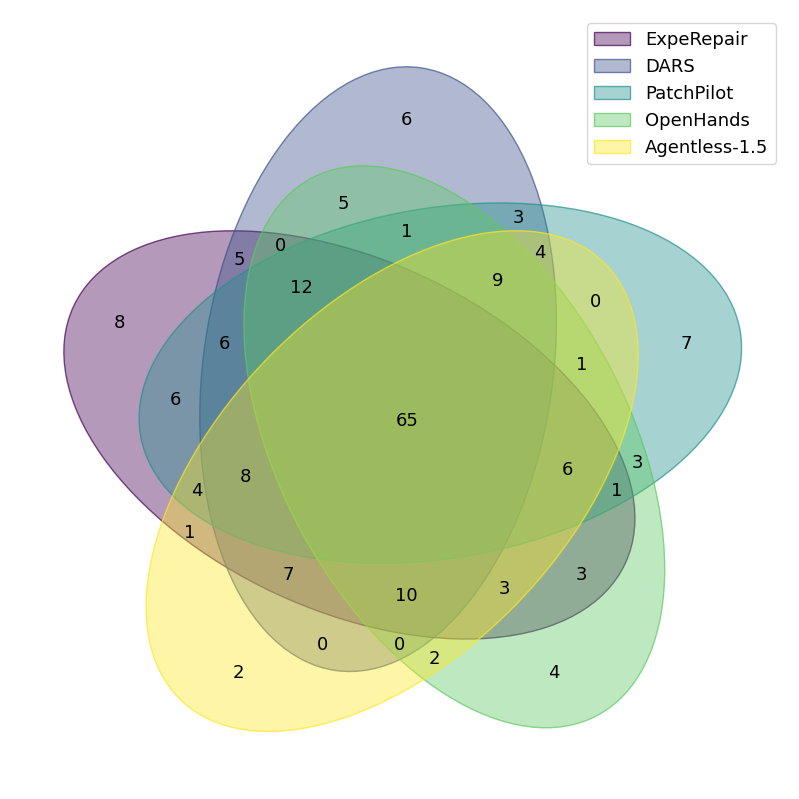}
        \vspace{-0.2cm}
        \caption{Intersection analysis.}
        \label{fig:rq1_1}
    \end{minipage}
    \hfill
    \begin{minipage}[b]{0.5\textwidth}
        \centering
        \includegraphics[width=\textwidth]{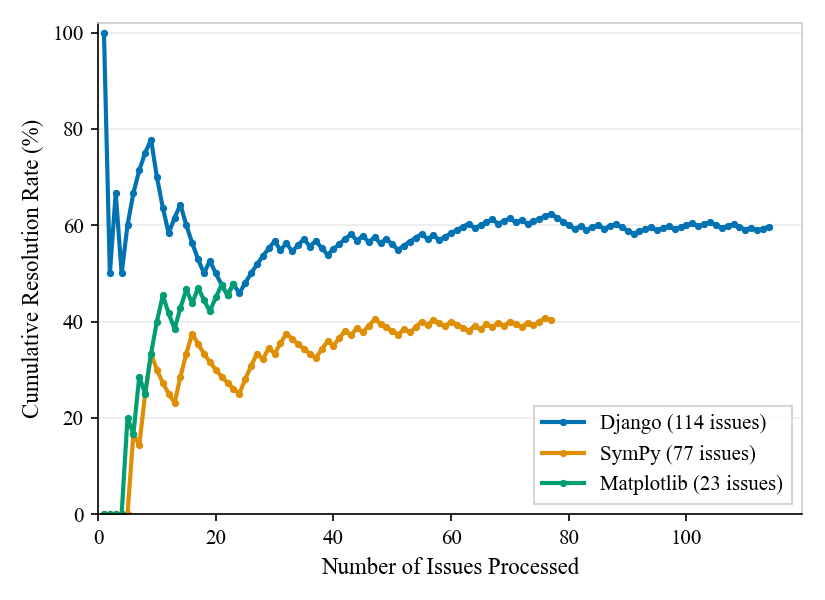}
        \vspace{-0.2cm}
        \caption{Cumulative resolution rates across repositories.}
        \label{fig:longitudinal}
    \end{minipage}
    \vspace{-0.35cm}
\end{figure}

As shown in Figure~\ref{fig:case_study}(b), this case examines a regression in Django 4.2 concerning the interaction between \texttt{QuerySet.only()} and \texttt{select\_related()} when applied to reverse \texttt{OneToOneField} relationships. In earlier versions (e.g., Django 4.1), restricting fields with \texttt{only()} correctly limited both the primary model and its related reverse model to the specified subset of fields. However, in Django 4.2, combining \texttt{only()} with \texttt{select\_related()} on a reverse \texttt{OneToOneField} fails to propagate field restrictions to the related model, resulting in all fields of the related model being included in the generated SQL. This deviation from expected behavior leads to unnecessary data retrieval, potentially impacting performance and memory usage.

In the initial attempt, the test agent tried to reproduce the issue without leveraging episodic memory. The generated script was ineffective due to an incomplete test setup: it did not establish a full Django application with proper \texttt{INSTALLED\_APPS}, lacked complete database table creation and test data population, and did not force query execution to generate SQL. Consequently, the queries either did not execute or did not reflect the field restrictions necessary to trigger the regression.

In a subsequent attempt, the agent accessed episodic memory to retrieve prior examples of testing scripts. The retrieved example correctly sets up a temporary Django application, creates the relevant models and test data, and executes queries in a controlled environment. Guided by these examples, the agent produced a fully functional test script that successfully reproduced the issue: when combining \texttt{select\_related()} with \texttt{only()}, all fields of the reverse \texttt{OneToOneField} model were included in the generated SQL, confirming the regression in Django 4.2.

\subsection{Intersection Analysis}

We further conduct an intersection analysis comparing the issues resolved by {\tool} against four leading baselines. Figure~\ref{fig:rq1_1} presents the complete overlap structure on SWE-Bench Lite.

As shown in Figure~\ref{fig:rq1_1}, {\tool} uniquely resolves 8 issues that no other open-source method fixes. Meanwhile, DARS and PatchPilot uniquely solve 6 and 7 issues, respectively, indicating that the repair capabilities of different methods are partially complementary rather than dominated by a single approach. In addition, 65 issues are solved by all methods, suggesting a shared ability to address relatively straightforward bugs.

We manually inspected the 8 issues uniquely fixed by {\tool} and found that six of their solutions follow recurring repair patterns derived from accumulated experience, which include: (i)~`add missing validation', (ii)~`generalize single-value to collection', and (iii)~`correct attribute or type resolution'. For example, in \texttt{django\_\_django-15320}, \texttt{Subquery.as\_sql()} produced invalid SQL because the query's \texttt{subquery} flag was unset. {\tool} correctly set \texttt{self.query.subquery = True} in the Subquery constructor by transferring the pattern of ensuring proper initialization from previous fixes, guiding the patch generation process.

At the same time, issues solved exclusively by DARS highlight a limitation of {\tool}: its effectiveness drops when relevant prior experience is unavailable. Four of the six DARS-only issues require novel solutions with few analogues in the repository, such as introducing new classes or overrides with custom behavior. Such cases favor DARS, whose structured branching and iterative refinement are better suited to exploratory reasoning than to analogy-based retrieval.

These findings suggest that {\tool} and the SOTA baselines exhibit complementary inductive biases, and that hybrid approaches leveraging their respective strengths offer a promising direction for future research.

\subsection{Longitudinal Study}

To assess whether {\tool}'s accumulated memory enhances repair performance, we conduct a longitudinal study to track how the cumulative resolution rate evolves as issues are processed. Since {\tool} maintains separate episodic and semantic memories per repository, we selected the three repositories in SWE-Bench Lite with the largest issue counts to ensure sufficient temporal depth for observing potential learning effects. Issues were processed in the chronological order provided by the dataset, and after each issue, we computed the cumulative resolution rate as the proportion of resolved issues among all issues processed up to that point.

As shown in Figure~\ref{fig:longitudinal}, all repositories exhibit high volatility during the early phase. Django's cumulative resolution rate fluctuates between 50\% and 100\% during the first 20 issues, whereas the resolution rates of SymPy and Matplotlib rise from zero with pronounced swings. After approximately 20 issues, the cumulative resolution rates begin to stabilize and gradually improve, suggesting that newly accumulated memory enhances coverage of recurring issue patterns, thereby improving overall performance and reducing variance. Django's cumulative resolution rate stabilizes around issue 80 at approximately 60\%, which may be attributable to diminishing returns from repeated memory and the increasing heterogeneity of the remaining issues.

\subsection{Limitation}
\label{sec:limitation}
While our approach improves test generation and patch generation by leveraging accumulated episodic and semantic memories, it currently places less emphasis on optimizing bug localization, which is also a critical factor in program repair performance. The primary challenge lies in the difficulty of verifying localization correctness. Unlike test generation and patch generation, whose correctness can be directly validated through execution outcomes (e.g., triggering issues or passing test cases), bug localization lacks an automated oracle. The correctness of a localization prediction cannot be reliably determined based on the success or failure of the subsequent patch. One possible solution is to selectively accumulate localization experiences only from successful repair attempts, assuming that a correct patch implies the localization was correct. However, this conservative strategy would result in memories containing only a narrow subset of successful localization cases, limiting both diversity and coverage. We leave the integration of reliable, memory-based bug localization into our framework as an important direction for future work.

\subsection{Threats to Validity}

The first threat to validity is the potential for data leakage. Since LLMs are often trained on open-source repositories, public benchmarks may inadvertently overlap with their training corpora. This raises the risk that evaluation results reflect prior exposure rather than genuine generalization. To mitigate this concern, we avoid long-standing APR benchmarks (e.g., Defects4J \cite{DBLP:conf/issta/JustJE14}), which are more likely to have been included in model training corpora. Instead, we adopt SWE-Bench, a more recent benchmark that is less susceptible to contamination and has gained broad adoption in recent studies \cite{swe-fixer, DARS, xia2024agentless}. Moreover, our evaluation emphasizes the reasoning process of models rather than their ability to recall existing solutions. Across experiments, our approach consistently outperforms strong baselines built on the same underlying models, suggesting that the observed gains stem from our methodology rather than memorization. In future work, we plan to extend our assessment to contamination-free datasets, such as live-updatable benchmarks \cite{DBLP:journals/corr/abs-2505-23419}, to further strengthen the robustness of our findings.

A second threat arises from the limited scope of our experiments, primarily constrained by computational resources. Although we evaluate four models from three different providers, our study still covers only a limited set of representative LLMs rather than the full spectrum of open- and closed-source architectures. This may introduce model-selection bias and limit the generalizability of our findings. To mitigate this concern, we draw on prior work in repository-level APR \cite{xia2024agentless}, which has shown the validity of evaluation on a limited number of models. Moreover, the consistent improvements observed across the evaluated models indicate that our approach is not tied to a particular model and is adaptable to diverse architectures.

The third threat concerns the generalizability of our approach beyond Python codebases. Our evaluation is limited to open-source projects implemented in Python, constraining our ability to demonstrate cross-language effectiveness. However, {\tool} is designed to be language-agnostic. It leverages experiential knowledge from test generation and bug repair patterns, rather than relying on language-specific syntax or semantics. This design suggests that our approach can extend to other programming languages, such as C\# and Ruby.

%% file: sec/related_work.tex
\section{Related Work}

\subsection{Memory-Enhanced LLM Agents}
To enhance LLM agents' ability to retain and exploit prior experiences, researchers have proposed a variety of memory-augmented architectures \cite{DBLP:journals/corr/abs-2502-12110, DBLP:conf/aaai/ZhongGGYW24, DBLP:journals/corr/abs-2311-08719, DBLP:conf/aaai/Zhao0XLLH24}.

Early explorations focused on human-inspired memory systems for long-term interactions \cite{DBLP:conf/aaai/ZhongGGYW24, DBLP:journals/corr/abs-2311-08719}. For instance, Think-in-Memory (TiM) \cite{DBLP:journals/corr/abs-2311-08719} introduced a two-phase paradigm consisting of pre-generation recall and post-generation memory refinement, allowing LLMs to evolve their memory representations without redundant reasoning steps. MemoryBank \cite{DBLP:conf/aaai/ZhongGGYW24} distinguished between short- and long-term memory to enable more natural human–AI communication. MemGPT \cite{DBLP:journals/corr/abs-2310-08560} adopted a hierarchical memory organization for extended context management. Collectively, these pioneering efforts highlighted the necessity of persistent memory systems, although their primary emphasis remained on dialogue continuity rather than specialized problem-solving tasks.

More recent frameworks move beyond conversational memory and aim to capture procedural knowledge derived from agent–environment interactions \cite{DBLP:journals/corr/abs-2504-19413, DBLP:journals/corr/abs-2502-12110, DBLP:conf/aaai/Zhao0XLLH24}. For example, ExpeL \cite{DBLP:conf/aaai/Zhao0XLLH24} extracts natural language insights and manages them via weighted voting (ADD, EDIT, UPVOTE, DOWNVOTE) for non-parametric learning, while AgentRR \cite{DBLP:journals/corr/abs-2505-17716} adopts a record-and-replay mechanism that logs environmental dynamics and decision trajectories. Building on these efforts, A-Mem \cite{DBLP:journals/corr/abs-2502-12110} introduces a Zettelkasten-inspired architecture with continuously linked notes for scalable retrieval, and Mem0 \cite{DBLP:journals/corr/abs-2504-19413} extracts key facts via a lightweight two-stage process, with its graph-based extension $\text{Mem0}^g$ supporting temporal and multi-hop reasoning. Complementing these, AriGraph \cite{anokhin2024arigraph} unifies semantic and episodic memory within a dynamic knowledge graph world model, allowing agents to reason over both abstract knowledge and concrete experiences, thereby enhancing planning and exploration in interactive environments.

The findings of previous work motivate the study presented in this paper. Our work differs from prior research in that it focuses on addressing the unique challenges of large-scale, collaborative software maintenance, bridging the gap between autonomous agent reasoning and practical program repair in real-world repositories.

\subsection{Repository-Level Automatic Program Repair}
Repository-level APR has attracted increasing attention from both academia and industry~\cite{chen2024large,bouzenia2024repairagent,zhang2024autocoderoverautonomousprogramimprovement,xia2024agentless,yang2024swe,DBLP:conf/icse/0009W0LRLW24}, motivated by the growing complexity of managing bug reports and issue tracking in collaborative platforms such as GitHub. With the rapid progress of LLMs in generative software engineering, LLM-based solutions have become the dominant paradigm for this problem space.

Early methods primarily relied on Retrieval-Augmented Generation (RAG) \cite{mansur2024ragfix}, which retrieves relevant code snippets to guide LLMs during patch generation \cite{jimenez2023swe}. More recently, agent-based frameworks treat LLMs as planning agents that iteratively interact with external tools to perform end-to-end issue resolution. Representative examples include SWE-agent \cite{yang2024swe}, which introduced a general agent–computer interface, together with Moatless \cite{moatless-tools}, OpenHands \cite{wang2024openhands}, AutoCodeRover \cite{zhang2024autocoderoverautonomousprogramimprovement}, and SpecRover \cite{DBLP:conf/icse/Ruan0R25}, which extend this paradigm with improved localization, multi-agent collaboration, and tighter tooling integration. SWE-Search \cite{swe-search} and DARS \cite{DBLP:journals/corr/abs-2503-14269} further enhance agent capabilities through better search and exploration strategies.

In parallel, procedural methods follow expert-designed, fixed workflows that decompose repair into structured phases (e.g., reproduction, localization, patching, validation), as exemplified by Agentless \cite{xia2024agentless} and PatchPilot \cite{li2025patchpilot}. Training-based pipelines such as SWE-Fixer \cite{swe-fixer} and MCTS-Refined CoT \cite{mcts-cot} further improve performance via continued training on real-world or synthetic repair data.

Although existing research has achieved promising results in repository-level APR, current approaches face a fundamental limitation: they typically treat each issue in isolation and lack systematic mechanisms to leverage historical repair experiences. To address this gap, {\tool} introduces a dual-memory system that continuously accumulates, organizes, and reuses past repair trajectories. By enabling dynamic prompt adaptation during inference, {\tool} improves repair effectiveness without relying on costly fine-tuning or complex agent orchestration.

%% file: sec/conclusion.tex
\section{Conclusion}
In real-world software development, developers continually accumulate knowledge from past debugging and repair activities, refining their skills and strategies over time. Inspired by this process, we present {\tool}, a novel LLM-based APR method that continuously accumulates and reuses historical repair experiences via a dual-memory system. By integrating episodic and semantic memories, {\tool} enables dynamic, context-aware prompt adaptation, enhancing both repair effectiveness and efficiency. We evaluate {\tool} on the SWE-Bench Lite and SWE-Bench Verified benchmarks, achieving 60.3\% pass@1 on SWE-Bench Lite and 74.6\% pass@1 on SWE-Bench Verified using Claude 4 Sonnet. These results demonstrate that leveraging past repair knowledge significantly improves performance on challenging program repair tasks.

\section{Data Availability}
The data and code necessary to replicate the results of this paper are publicly available at \url{https://github.com/ExpeRepair/ExpeRepair}.

\section*{Acknowledgments}
We sincerely appreciate anonymous reviewers for their constructive and insightful suggestions for improving this manuscript.
This work was supported by the National Natural Science Foundation of China Grant No.62232016, Key Project of ISCAS (Grant no. ISCAS-ZD-202401), 
Basic Research Program of ISCAS (Grant no. ISCAS-JCZD-202304), Innovation Team 2024 ISCAS (No. 2024-66).